\documentclass[pre,groupedaddress,showkeys,showpacs,twocolumn]{revtex4}
\usepackage{amsmath}
\usepackage{graphics}
\usepackage{graphicx}
\usepackage{amsfonts}
\usepackage{amssymb}
\usepackage{dcolumn}
\usepackage{bm}
\usepackage{natbib}
\begin{document}
\title{Non Amontons-Coulomb local friction law of randomly rough contact interfaces with rubber}
\author{Danh Toan Nguyen$^{1,3}$}
\author{Elie Wandersman$^{2}$}
\author{Alexis Prevost$^{2}$}
\author{Yohan Le Chenadec$^{3}$}
\author{Christian Fr\'etigny$^{1}$}
\author{Antoine Chateauminoisy$^{1}$}
\email[]{antoine.chateauminois@espci.fr}
\affiliation{1. Soft Matter Science and Engineering Laboratory (SIMM), UMR CNRS
7615,
Ecole Sup\'erieure de Physique et Chimie Industrielles (ESPCI), Universit\'e Pierre et Marie Curie, Paris (UPMC), France\\
2. CNRS / UPMC Univ Paris 06, FRE 3231, Laboratoire Jean Perrin, F-75005, Paris, France\\
3. Manufacture Fran\c{c}aise des Pneumatiques Michelin, 63040 Cedex 9, Clermont-Ferrand, France\\}
%
%
%
%
\begin{abstract}
We report on measurements of the local friction law at a multi-contact interface formed between a smooth rubber and statistically rough glass lenses, under steady state friction. Using contact imaging, surface displacements are measured, and inverted to extract both distributions of frictional shear stress and contact pressure with a spatial resolution of about 10~$\mu$m. For a glass surface whose topography is self-affine with a Gaussian height asperity distribution, the local frictional shear stress is found to vary strongly sub-linearly with the local contact pressure over the whole investigated pressure range. Such sub-linear behavior is also evidenced for a surface with a non Gaussian height asperity distribution, demonstrating that, for such multi-contact interfaces, Amontons-Coulomb's friction law does not prevail at the local scale.
\end{abstract}
\maketitle
\section{Introduction}
Friction is one of the long standing problems in physics which still remains partially unsolved. Similarly to adhesive contact problems, friction couples mechanical properties of the materials in contact, roughness and physicochemical characteristics of their surfaces. To incorporate such intricate effects in a description of friction, one needs to postulate a local constitutive law indicating how shear stresses depend on normal stresses at the interface. For macroscopic contacts, Bowden and Tabor~\cite{Bowden1958} and later Greenwood and Williamson~\cite{greenwood1966} were the first to recognize the crucial contribution of surface roughness in the derivation of such constitutive laws. Their approach to friction is based on the observation that, due to the distribution of asperities heights on the surface, contact between two macroscopic solids is usually made up of a myriad of micro-contacts. The real area of contact is thus much smaller than the macroscopic apparent one. As a result,
friction of multi-contact interfaces combines multiple length scales. At the scale of a single asperity, frictional energy dissipation involves poorly understood physicochemical processes occurring at the intimate contact between surfaces, like adsorption or entanglement/distanglement mechanisms for instance~\cite{bureau2004,drummond2007}, as well as viscoelastic or plastic deformations of the asperities~\cite{greenwood1958,grosch1963a}. At the macroscopic scale, \textit{i.e.} the size of the contact, friction processes involve the collective contact mechanics of a statistical set of asperities whose sizes are often distributed over orders of magnitude. Several models were proposed for evaluating the area of real contact and its dependence on the normal load, often based on a spectral description of surface roughness~\cite{Archard_1957,campana2007,campana2008,greenwood1966,persson2001}. One of the key issues of these models is to incorporate in a realistic way the effects of 
adhesion and materials properties such as plasticity and viscoelasticity on the formation of the actual contact area under sliding conditions.\\
\indent This concept of real contact area is central to sliding situations where the overall friction force is usually assumed to be the sum of the shear resistance of individual micro-contacts. As a crude assumption, the friction force can be considered as the product of the actual contact area by a constant shear stress which embeds all dissipative mechanisms occurring at the scale of micro-contacts. This idea forms the basis of the Bowden and Tabor model~\cite{Bowden1958} which was later enriched to account for rate dependence and aging effects on friction~\cite{bureau2002,ruina1983}. As reviewed in ~\cite{baumberger2006}, it remains the current framework for the description of solid friction at multi-contact interfaces. 
\indent Experimentally, validation of such models mostly relies on measurements of the friction force and its dependence on normal load and sliding velocity. Unfortunately, friction force is an average of local frictional properties which makes the validation of local friction laws, and, \textit{a fortiori}, of the proposed models rather indirect. Knowledge of a local constitutive friction law is however relevant to many contact mechanics models where local friction at contact interfaces is often postulated to obey locally Amontons-Coulomb's friction law~\cite{Johnson1985}. It also remains crucial in our understanding of induced non-linear friction force fluctuations which are exhibited for instance in tactile perception~\cite{wandersman2011}.\\
\indent In this Letter, we take advantage of a previously developed experimental method~\cite{chateauminois2008,nguyen2011} for the determination of shear stress and contact pressure distributions within contacts to address the problem of a frictional interface between a smooth silicone rubber and a rigid randomly rough surface. The approach is based on the measurement of the displacement field at the surface of the rubber substrate which, after inversion, provides the corresponding distributions of both local contact pressure and frictional shear stress within the contact. The method is first applied to a frictional interface with a self-affine fractal roughness and Gaussian height asperity distribution, allowing us to measure a local friction law at length scales much smaller than the size of the contact. Its relationship with the macroscopic friction law is also discussed. The method is then applied to a non-Gaussian surface, allowing to probe how the local friction law is affected by a change of 
topography.\\
\section{Experimental details}
\begin{figure}[ht!]
	\centering
	\includegraphics[width=\columnwidth]{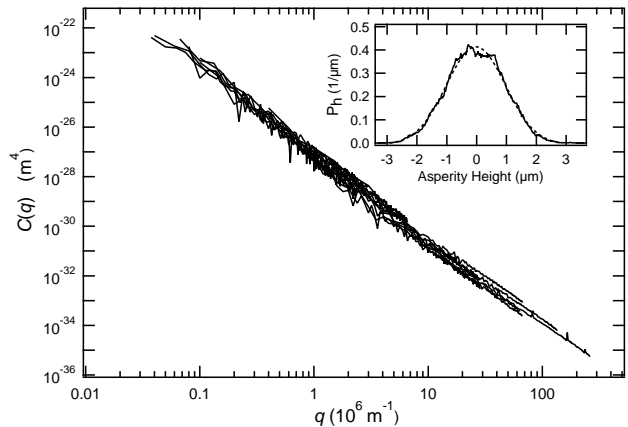}
	\caption{$C(q)$ for the sand blasted glass lens. Overlapping segments correspond to different AFM measurements on different regions of interest. $C(q)$ follows a power law in the range $0.1-100\:10^6\:m^{-1}$. Inset: Height distribution for the present sample (solid line). The dotted line is a Gaussian fit.}
	\label{fig:PSD_rough_lens}
\end{figure}

\indent A commercially available  transparent Poly(DiMethylSiloxane) silicone (PDMS Sylgard 184, Dow Corning, Midland, MI) is used as an elastomer substrate. In order to monitor contact induced surface displacements, a square network of small cylindrical holes (diameter 8 $\mu$m, depth 11 $\mu$m and center-to-center spacing 400 $\mu m$) is stamped on the PDMS surface by means of standard soft lithography techniques. Once imaged in transmission with a white light, the pattern appears as a network of dark spots which are easily detected using image analysis. Full details regarding the design and fabrication of PDMS substrates are provided in ~\cite{nguyen2011}. Their dimensions ($15~\times~60~\times~60$~mm$^{3}$) ensure that semi-infinite contact conditions are met during friction experiments (\textit{i.e.} the ratio of the substrate thickness to the contact radius is larger than 10 \cite{gacoin2006}). Before use, PDMS substrates are thoroughly washed with isopropanol and subsequently dried in a vacuum 
chamber kept at low pressure.
\indent Millimeter sized contacts are achieved between the PDMS substrate and plano-convex BK7 glass lenses of radius of curvature 5.2~mm (Melles Griot, France). Their surface are rendered microscopically rough using sand blasting (average grains size of 60~$\mu$m). The resulting topography has been characterized using AFM measurements over increasingly large regions of interest, from $0.5~\times~0.5~\mu m^2$ up to $80~\times~80~\mu m^2$. This allowed to probe its roughness at multiple length scales $\lambda$ from $50\ \mu m$ down to a few nanometers, and compute its height distribution and height Power Spectrum Density (PSD) $C(q)$ \cite{persson2001}, where $q=2 \pi/\lambda$ is the wave vector. The height distribution is found to be Gaussian with a standard deviation $\sigma= 1.40 \pm 0.01~\mu$m (Fig.~\ref{fig:PSD_rough_lens}, inset), and $C(q)$ follows a power law at all $q$, characteristic of self-affine fractal surfaces (Fig.~\ref{fig:PSD_rough_lens}). Fitting $C(q)$ with its expected functional form $C(
q)\propto q^{-2(H+1)}$ for this type of topography yields a Hurst exponent $H=0.74$ and a fractal dimension $D_{f}=3-H=2.26$. This sand-blasted glass surface is sometimes referred to as a Gaussian surface. \\
Depending on the investigated normal load range, friction experiments are performed using two different custom-built setups designed respectively for high normal loads $P$ (1 to 17~N) and low $P$ (0.02 to 2~N). Both setups, which are described elsewhere~(respectively \cite{chateauminois2008} and \cite{prevost2013}), are operated at constant $P$ and constant sliding velocity. The PDMS substrate is displaced with respect to the fixed glass lens by means of a linear translation stage while the lateral load $Q$ is continuously recorded either using a load transducer for the high load setup, or using a combination of a shear cantilever and a capacitive displacement sensor for the low load setup. For all experiments, smooth friction is achieved with no evidence of stick-slip instabilities nor detachment waves \cite{rubinstein2007,rubinstein2009}. Experiments carried out between 0.01 and 10~mm~s$^{-1}$ did not reveal any strong changes in the frictional behavior and thus, only results obtained at the intermediate 
velocity of 0.5~mm~s$^{-1}$ are reported in the present paper\footnote{This specific value was chosen as it falls within both accessible velocity ranges for both setups. Indeed, the high load setup can be operated at a maximum driving velocity of 10~mm~s$^{-1}$, while the low load setup at 1~mm~s$^{-1}$.}. During steady state friction, images of the deformed contact zone are continuously recorded through the transparent PDMS substrate using a zoom lens and a camera.  The system is configured to a frame size of $1024~\times~1024$ pixels$^2$ with 8 bits resolution. For each image, positions of the markers are detected with a sub-pixel resolution using a dedicated image processing software. Accumulation of data from a set of about 400 successive images at a maximum frame rate of 24~Hz results in a well sampled lateral displacement field with a spatial resolution of $\sim 10\: \mu m$, which is much larger than the markers' spacing ($400\: \mu m$). The accuracy in the measurement of the lateral displacements is 
better than 1~$\mu$m.\\
\indent Surface displacement fields are inverted to extract the corresponding contact stress distribution. As detailed in~\cite{nguyen2011}, a three dimensional Finite Element (FE) inversion procedure has been developed which takes into account the non linearities arising from the large strains (up to $\approx 0.4$) which are often induced at the edges of the contact, in particular at high normal loads. The principle of the approach is to apply the surface displacement field as a boundary condition at the upper surface of a meshed body representing the rubber substrate and to compute the corresponding stress distribution under the assumption of a Neo-Hookean behavior of the PDMS material~\cite{nguyen2011}. In addition to the measured lateral displacement field, the vertical displacements of the PDMS surface within the contact area are also used as a boundary condition in order to compute the contact pressure distribution. Vertical displacements are not measured locally within the contact but they 
are deduced using both the radius of curvature of the glass lens and the measured indentation depth under steady state sliding. In other words, a nominal vertical displacement field is used in the inversion which does not include micrometer scale variations due to the surface roughness. Such an approach is expected not to affect the pressure field if asperities heights remain low as compared to the nominal vertical displacement. Such an assumption is likely to be valid except very close to the contact edge or at very low applied normal loads.
\indent After the numerical inversion calculation, the local contact pressure and frictional shear stress are determined from a projection of the stress tensor in a local cartesian coordinate system whose orientation is defined from the normal to the lens surface and from the actual sliding direction. The inversion procedure thus takes into account the contact geometry together with the measured sliding path trajectories.\\
\section{Contact pressure and shear stress fields}
\begin{figure}[ht!]
	\centering
	\includegraphics[width=\columnwidth]{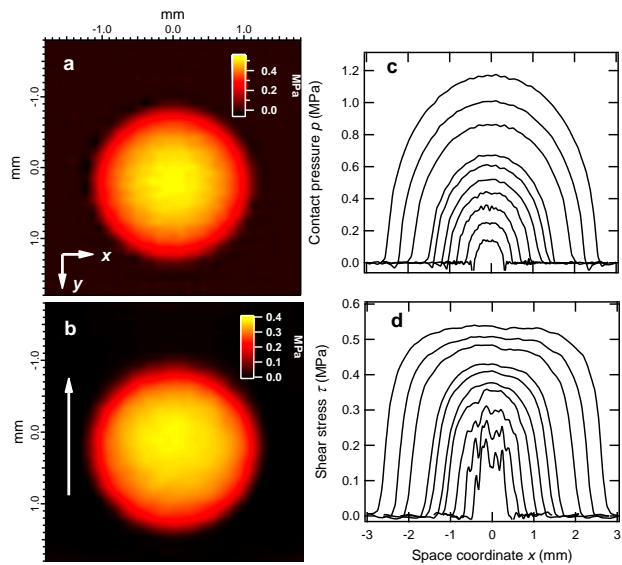}
	\caption{2D maps of the pressure (a) and shear stress (b) distributions (in MPa) within a frictional contact between a PDMS substrate and the sand blasted lens, at $P= 1.6$~N. The white arrow in (a) shows the direction of sliding. (c) Profiles of contact pressure and (d) frictional shear stress taken across the contact at $y=0.2\:mm$ and perpendicular to the sliding direction for different $P$. From bottom to top: $P=$ 0.06, 0.2, 0.5, 1.0, 1.6, 2.5, 3.5, 7.3, 11.2 and 17.0 N.}
	\label{fig:field_rough}
\end{figure}

\begin{figure*}[ht!]
	\centering
	\includegraphics[width=1.5\columnwidth]{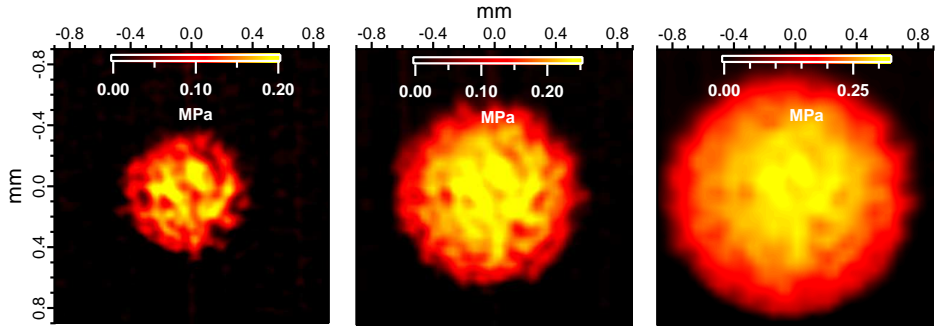}
	\caption{2D maps of the shear stress fields $\tau(x,y)$ at low contact pressures. $P$ is from left to right, 0.05, 0.2 and 0.5 N.}
	\label{fig:field_rough_lowp}
\end{figure*}

\indent Figures~\ref{fig:field_rough}a and \ref{fig:field_rough}b show an example of the contact pressure and shear stress spatial distributions, respectively $p(x,y)$ and $\tau(x,y)$, which are measured in steady sliding with the Gaussian rough surface. In what follows, it should be kept in mind that the reported stress data correspond to spatially averaged values over an area of about 10~$\mu$m$^2$, determined by the spatial resolution of the displacement measurement. Owing to the self affine fractal nature of the investigated rough surface, there are still many asperities in contact at this scale. Measured values of the frictional shear stress thus represents a statistical average which encompasses all roughness length scales up to about 10~$\mu$m. In Fig.~\ref{fig:field_rough}b, the frictional shear stress distribution shows a shape similar to that of the contact pressure with a maximum at the centre of the contact (Fig.~\ref{fig:field_rough}a). This correlation is further evidenced in Figs.~\ref{fig:field_rough}c and \ref{fig:field_rough}d where sections of the shear stress and contact pressure fields taken across the contact area and perpendicular to the sliding direction are reported for increasing normal loads. Contact pressure profiles show a bell-shaped Hertz-like distribution which is expected from the prescribed spherical distribution of vertical displacements within the contact area. However, the measured pressure distribution takes into account the non linearities arising from finite strain together with mechanical coupling between normal and lateral stresses as previously reported~\cite{nguyen2011}. Similar frictional shear stress profiles are also obtained for an increasing $P$ but with some evidence of a saturation at high contact pressure. Such a dependence of the frictional shear stress on the applied contact pressure reflects the multi-contact nature of the interface. As the local contact pressure is increased, the number of micro-contacts grows, thus enhancing local frictional shear 
stresses. As mentioned in previous studies~\cite{nguyen2011,nguyen2013}, such pressure dependence is not observed within frictional contacts between PDMS and a smooth glass lens where intimate contact is achieved.\\
At low normal loads ($P \leq 0.5~$N), stress fluctuations are clearly present in the shear stress profiles (Fig.~\ref{fig:field_rough}d). Looking at 2D spatial maps of the stress fields for these loads (Fig.~\ref{fig:field_rough_lowp}) reveals that these fluctuations are distributed spatially over length scales of the order of a few tens of micrometers. Close examination of the shear stress fields measured for three different $P$ actually shows that features of the stress field at a given location within the contact remain at the same location when $P$ is increased. The observed variations of the shear stress at small $P$ thus likely reflect local changes in the contact stress distribution which are induced by details of the topography of the rough lens at these length scales. This result thus demonstrates the ability of displacement fields measurements and inversion procedure to probe spatial fluctuations in the shear stress distribution down to a few tens of micrometers. At higher $P$, stress spatial 
variations are blurred out, most likely as a result of an increasing intimate contact between surfaces.\\
\section{Local friction law}
\begin{figure}[ht!]
	\centering
	\includegraphics[width=0.8\columnwidth]{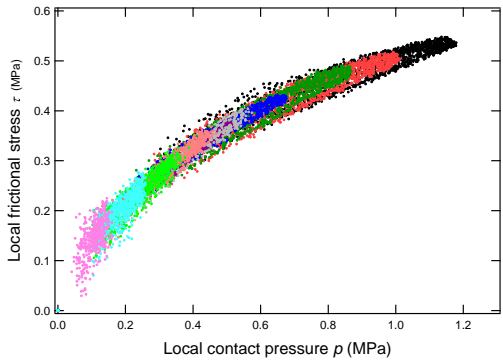}
	\caption{Local shear stress $\tau$ as a function of local contact pressure $p$ for a rough Gaussian contact. Colors denote different friction experiments at increasing $P$ from 55~mN to 17~N. Each data point corresponds to a given location in the contact.}
	\label{fig:local_friction_law}
\end{figure}

\begin{figure}[ht!]
	\centering
	\includegraphics[width=0.7\columnwidth]{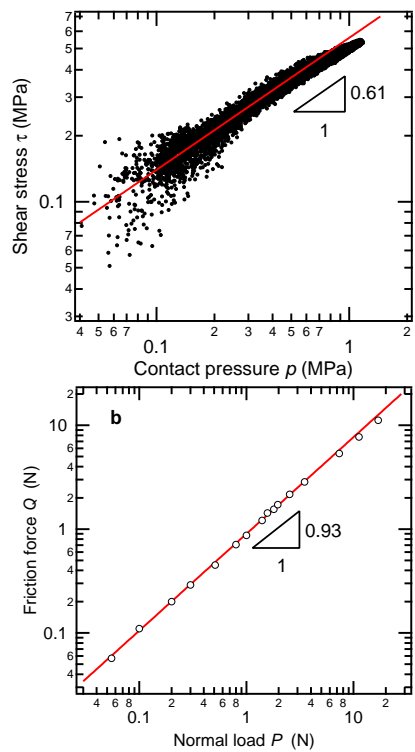}
	\caption{(a) Log-log plot of $\tau$ \textit{vs.} $p$ (data of Fig.~\ref{fig:local_friction_law}). (b) Log-log plot of $Q$ \textit{vs.} $P$. On both plots, solid lines are power law fits.}
	\label{fig:powerlawfit}
\end{figure}

\begin{figure}[ht!]
	\centering
	\includegraphics[width=\columnwidth]{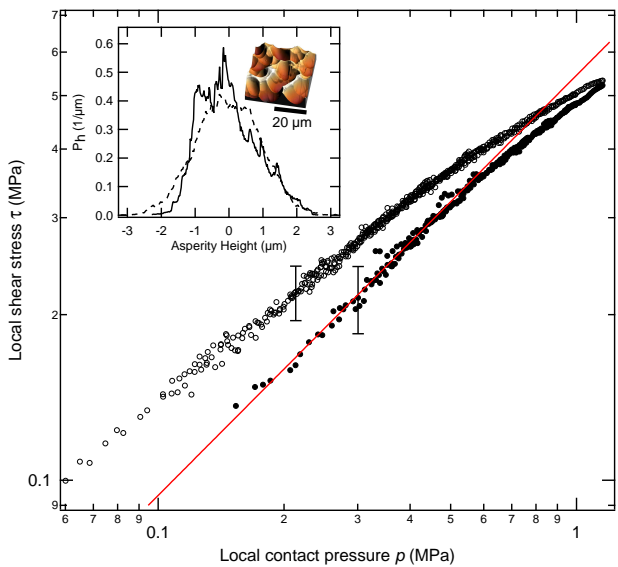}
	\caption{Averaged local friction law of both sand blasted glass lens with Gaussian roughness ($\circ$) and sand blasted and etched glass surface with a non Gaussian, cusp-like topography (\textbullet). Vertical bars give the extent of the non-averaged data for both sets. The solid line corresponds to a power law fit of the cusp data. Inset: Non Gaussian height distribution $P_h$ of the sand blasted and etched glass surface as measured using AFM (solid line). A 3D rendering of the surface, obtained from AFM measurements, is shown in the upper corner. For comparison, the Gaussian height distribution of the sand blasted surface is shown (dotted line).}
	\label{fig:cusps}
\end{figure}

\indent We now examine more closely the relationship between contact pressure and frictional stress, \textit{i.e.} the local friction law. The existence of a well defined relationship between local shear stress $\tau$ and contact pressure $p$ would imply that all data points obtained at different $P$ and different positions $(x,y)$ within the contact should merge onto a single curve, when reported in a $(\tau,p)$ plane. Such a master curve is indeed obtained as clearly shown on Fig.~\ref{fig:local_friction_law}. In this figure, each color corresponds to a different $P$ and each data point to a given location within the contact. The local contact pressure profile is close to a Hertzian one (Fig.~\ref{fig:field_rough}c), but does not take into account roughness induced deviations which were predicted theoretically by Greenwood and Tripp~\cite{greenwood1967}. Such deviations include at low nominal contact pressure both a decrease of the maximum $p$ at the center of the contact and the existence of a tail in the 
pressure distribution at the contact edges \cite{scheibert2008}. As a result of such effects, one should especially expect systematic deviations from the master curve of data points obtained in the low pressure range (\textit{i.e.} in the vicinity of the contact edges) for each of the considered $P$. This is not observed in Fig.~\ref{fig:local_friction_law} which tends to indicate that deviations from Hertz pressure distribution, induced by surface roughness, are not significant in our analysis.\\
\indent The obtained local friction law is markedly sub-linear over the whole investigated contact pressure range. If one makes the assumption that the shear stress is increasing with the local density of micro-contacts, the observed sub-linear response should reflect the fact that the proportion of area in contact progressively saturates when contact pressure is increased. Saturation of the contact area at all length scales should eventually result in a constant, pressure independent frictional stress. Results shown in Fig.~\ref{fig:local_friction_law} indicate that such a saturation would occur at contact pressures close to or higher than the Young's modulus of the PDMS substrate ($E=$~3~MPa).\\
The measured local friction law can be fitted from the lowest pressures experimentally available up to $p=0.5$~MPa by a power law, $\tau(x,y)=\beta p(x,y)^m$ with $\beta=0.560\pm 0.003$ and $m=0.61 \pm 0.03$ (Fig.~\ref{fig:powerlawfit}a). For the rough contact interface considered here, such a local friction law differs significantly from Bowden and Tabor's classical expression ~\cite{Bowden1958}, \textit{i.e.} $\tau=\tau_0+\alpha p$, since the so-called adhesive term $\tau_0$ is negligible and that the pressure dependent term is markedly non linear. Assuming that $p$ follows a Hertzian profile, integrating $\tau(x,y)$ over the contact area yields the total friction force $Q$ which is found to scale with $P$ as $Q \propto P^{\gamma}$ with $\gamma=(m+2)/3$. This power law dependence is effectively obtained from friction force measurements as shown in Fig.~\ref{fig:powerlawfit}b. The experimental value of the exponent ($0.93 \pm 0.01$) is very close to that derived from the integration of the local friction 
law,
$(m+2)/3=0.87~\pm 0.01$. Interestingly, the same functional form $\tau=\beta p^m$ was actually postulated by some of us~\cite{wandersman2011} in a previous study involving a soft PDMS sphere sliding against a rough rigid plane with a similar roughness as the one used in the present study. The set of parameters $(\beta,m)$ were deduced from the measured $Q$ versus $P$ relationships using the exact same derivation. Although both systems are in essence different, an exponent $0.87~\pm~0.03$ was found for $Q$ versus $P$ curves, yielding an exponent $m=0.63$ nearly equal to the one measured with the current data. As stated in the introduction, friction of rough multi-contact interfaces involves intricate aspects related to the determination of the real contact area and energy dissipation mechanisms at the scale of single asperities. A simple approach based on Greenwood and Williamson rough contact model with the assumption of a constant interfacial shear stress and a Gaussian asperity height distribution would yield an Amontons-Coulomb local friction law at a mesoscopic length scale. The measured sub-linear, non Amontons-Coulomb, friction law may arise from a combination of the progressive saturation of the real contact area at high loads and of possible elastic interactions between neighboring asperities. To our knowledge, no current contact mechanics model provides the derivation of such a local friction law preventing any further discussion of the physical meaning of both $m$ and $\beta$ and their dependence on surface properties.\\
In order to assess the sensitivity of the local friction law to the details of surface roughness, a different surface topography was produced by a chemical etching of the sand blasted glass surface in hydrofluoric acid. As detailed in~\cite{spierings1993}, etching silicate glass surfaces with blasting induced micro-flaws results in a surface containing small cusps. Such a structure is shown in the inset of Fig.~\ref{fig:cusps} together with the height distribution profile showing the non Gaussian nature of the rough surface. In the same figure, it can be seen that the cusp-like surface also yields a power law dependence of the local shear stress on the contact pressure with an exponent $m=0.67 \pm 0.06$, comparable to the one obtained with the Gaussian surface. Such a weak dependence of the exponent on roughness was also evidenced using macroscopic measurements ($Q$ versus $P$) in \cite{wandersman2011}. The main difference rather lies in the magnitude of the prefactor $\beta=0.45 \pm 0.02$ which is reduced  
for the cusp-like surface. Under the classical assumption that the local shear stress can be described as the product of the actual contact area by an average shear stress embedding all the dissipative mechanisms occurring at micro-asperity scale, this difference could potentially arise from two effects. The first one is obviously a reduction of the proportion of area in contact for a given contact pressure in the case of the cusp-like surface. The second effect at play could be a reduction in the extent of frictional energy dissipation at the scale of the asperity as a result, for example, of a change in viscoelastic losses involved in surface deformation at micro-asperity scale. A discussion of these effects would however require a detailed contact mechanics analysis of the rough surfaces which is beyond the scope of the present paper.
\section{Conclusion}
The local friction law of a rubber surface sliding against randomly rough rigid surfaces has been determined from a measurement of the surface displacement field. Measured contacts stresses being resolved down to a length scale of about 10~$\mu$m, they reflect the local frictional properties of the multi-contact interface. The local friction law exhibits a strongly non Amontons-Coulomb, sub-linear dependence on contact pressure. These features are preserved when the topography of the rough surface is changed from Gaussian to non Gaussian which tends to support the generality of the observations. These results question the validity of Amontons-Coulomb's law hypothesis embedded in most rough contact friction models. More generally, the determination of such local friction laws should serve as a basis for the validation of theoretical rough contacts models. We have also shown that our analysis is able to resolve shear stress fluctuations which are induced by the distribution of asperities 
size at length scales of the order of a few tens of micrometers. A statistical analysis should interestingly show some correlation between the features of these shear stress variations and roughness parameters. It would, however, deserve an extended set of experiments where shear stress fields are measured for different realizations of the statistically rough surface.\\
\begin{acknowledgments}
	This study was partially funded by ANR (DYNALO project NT09-499845). We thank B. Bresson for the AFM measurements and are indebted to S. Roux for stimulating discussions. We also thank F. Monti for helping us with the chemical etching of the glass surfaces.
\end{acknowledgments}
\bibliographystyle{unsrt}
%


\begin{thebibliography}{10}
	
	\bibitem{Bowden1958}
	F.P. Bowden and Tabor D.
	\newblock {\em The Friction and Lubrication of Solids}.
	\newblock Clarendon Press, Oxford, 1958.
	
	\bibitem{greenwood1966}
	JA~Greenwood and JBP Williamson.
	\newblock Contact of nominally flat surfaces.
	\newblock {\em Proceedings of the Royal Society of London. Series A.
		Mathematical and Physical Sciences}, 295(1442):300--319, 1966.
	
	\bibitem{bureau2004}
	L.~Bureau and L.~Leger.
	\newblock Sliding friction at a rubber/brush interface.
	\newblock {\em Langmuir}, 20:4523, 2004.
	
	\bibitem{drummond2007}
	C.~Drummond, J.~Rodr{\'{\i}}guez-Hern{\'{a}}ndez, S.~Lecommandoux, and
	P.~Richetti.
	\newblock Boundary lubricant films under shear: effect of roughness and
	adhesion.
	\newblock {\em Journal of Chemical Physics}, 126:Art 184906, 2007.
	
	\bibitem{greenwood1958}
	J.A. Greenwood and D.~Tabor.
	\newblock The friction of hard sliders on lubricated rubber: the importance of
	deformation losses.
	\newblock {\em Proceedings of the Physical Society}, 71:989--1001, 1958.
	
	\bibitem{grosch1963a}
	A.K. Grosch.
	\newblock The relation between the friction and visco-elastic properties of
	rubber.
	\newblock {\em Proceedings of the Royal Society of London. Series A.
		Mathematical and Physical Sciences}, 274(1356):21--39, 1963.
	
	\bibitem{campana2007}
	C.~Campana and M.~Muser.
	\newblock Contact mechanics of real versus randomly rough surfaces: A green's
	function molecular dynamics study.
	\newblock {\em Europhysics}, 77:38005, 2007.
	
	\bibitem{campana2008}
	C.~Campana, M.H. Muser, and M.O. Robbins.
	\newblock Elastic contact between self-affine surfaces: comparison of numerical
	stress and contact correlation functions with analytic predictions.
	\newblock {\em Journal of Physics-Condensed Matter}, 20:354013, 2008.
	
	\bibitem{persson2001}
	B.N.J. Persson.
	\newblock Theory of rubber friction and contact mechanics.
	\newblock {\em Journal of Chemical Physics.}, 115(8):3840--3861, 2001.
	
	\bibitem{bureau2002}
	L~Bureau, T.~Baumberger, and C.~Caroli.
	\newblock Rheological aging and rejuvenation in solid friction contacts.
	\newblock {\em European Physical Journal E}, 8:331--337, 2002.
	
	\bibitem{ruina1983}
	A.~Ruina.
	\newblock Slip instability and state variable friction laws.
	\newblock {\em Journal of Geophysical Research}, 88:359--370, 1983.
	
	\bibitem{baumberger2006}
	T.~Baumberger and C.~Caroli.
	\newblock Solid friction from stick-slip down to pinning and aging.
	\newblock {\em Advances in Physics}, 55:279--348, 2006.
	
	\bibitem{wandersman2011}
	E.~Wandersman, R.~Candelier, G.~Debregeas, and A.~Prevost.
	\newblock Texture-induced modulations of friction force: The fingerprint
	effect.
	\newblock {\em Physical Review Letters}, 107:164301, 2011.
	
	\bibitem{chateauminois2008}
	A.~Chateauminois and C.~Fretigny.
	\newblock Local friction at a sliding interface between an elastomer and a
	rigid spherical probe.
	\newblock {\em European Physical Journal E}, 27(2):221--227, oct 2008.
	
	\bibitem{nguyen2011}
	D.T. Nguyen, P.~Paolino, M-C. Audry, A.~Chateauminois, C.~Fr{\'{e}}tigny, Y.~Le
	Chenadec, M.~Portigliatti, and E.~Barthel.
	\newblock Surface pressure and shear stress field within a frictional contact
	on rubber.
	\newblock {\em Journal of Adhesion}, 87:235--250, 2011.
	
	\bibitem{prevost2013}
	A.~Prevost, J.~Scheibert, and G.~Debr\'egeas.
	\newblock Probing the micromechanics of a multi-contact interface at the onset
	of frictional sliding.
	\newblock {\em European Physical E}, 36:13017, 2013.
	
	\bibitem{rubinstein2007}
	S.~M. Rubinstein, G.~Cohen, and J.~Fineberg.
	\newblock Dynamics of precursors to frictional sliding.
	\newblock {\em Physical Review Letters}, 98(22), Jun 1 2007.
	
	\bibitem{rubinstein2009}
	S.M. Rubinstein, G.~Cohen, and J.~Fineberg.
	\newblock Visualizing stick-slip: experimental observations of processes
	governin the nucleation of frictional sliding.
	\newblock {\em Journal of Physics D: Applied Physics}, 42:214016, 2009.
	
	\bibitem{nguyen2013}
	D.~T. Nguyen, S.~Ramakrishna, C.~Fretigny, N.~D. Spencer, Y.~Le Chenadec, and
	A.~Chateauminois.
	\newblock Friction of rubber with surfaces patterned with rigid spherical
	asperities.
	\newblock {\em Tribology Letters}, 49(1):135--144, January 2013.
	
	\bibitem{greenwood1967}
	Jim~A Greenwood and J~Hl Tripp.
	\newblock The elastic contact of rough spheres.
	\newblock {\em Journal of Applied Mechanics}, 34:153, 1967.
	
	\bibitem{scheibert2008}
	J.~Scheibert.
	\newblock {\em M\'ecanique du contact aux \'echelles m\'esoscopiques.}
	\newblock Sciences M\'ecaniques et Physiques. Edilivres, 2008.
	
	\bibitem{spierings1993}
	G.A.C.M. Spierings.
	\newblock Wetchemical etching of silica glglass in hydrofluoric acid based
	solutions.
	\newblock {\em Journal of Material Science}, 28:6261--6273, 1993.
	
\end{thebibliography}
%
%
\end{document}